\documentclass[a4paper,11pt]{article}
\pdfoutput=1 

\usepackage{jinstpub} 

\usepackage{caption}
\usepackage{subcaption}

\usepackage{siunitx}

\usepackage{lineno}

\title{\boldmath Operating the GridPix detector with helium-isobutane gas mixtures for a high-precision, low-mass Time Projection Chamber}

\author[a,b]{G.~Cavoto,}
\author[c]{C.~Dutsov,}
\author[d]{M.~Gruber,}
\author[c]{M.~Hildebrandt,}
\author[c,e]{T.~D.~Hume,}
\author[d]{J.~Kaminski,}
\author[f]{F.~Neuhaus,}
\author[c,g,h]{A.~Papa,}
\author[a,1]{F.~Renga,\note{Corresponding author.}}
\author[c]{P.~Schmidt-Wellenburg,}
\author[f]{M.~Schott,}
\author[b,g]{B.~Vitali,}
\author[a,b]{and C.~Voena}

\affiliation[a]{Istituto Nazionale di Fisica Nucleare, Sez.~di Roma, P.le A.~Moro 2, 00185 Roma (Italy)}
\affiliation[b]{Dipartimento di Fisica dell'Università di Roma “La Sapienza”, Piazzale Aldo Moro 2, 00185 Roma (Italy)}
\affiliation[c]{Paul Scherrer Institut, Forschungsstrasse 111, 5232 Villigen (Switzerland)}
\affiliation[d]{Physikalisches Institut der Universit\"at Bonn, Nu{\ss}allee 12, 53115 Bonn (Germany)}
\affiliation[e]{Swiss Federal Institute of Technology ETH, Rämistrasse 101, 8092 Zürich (Switzerland)}
\affiliation[f]{Institut f\"ur Physik, Universit\"at Mainz, Staudingerweg 7, 55128 Mainz (Germany)}
\affiliation[g]{Istituto Nazionale di Fisica Nucleare, Sez.~di Pisa, Largo B. Pontecorvo 3, 56127 Pisa (Italy)}
\affiliation[h]{Dipartimento di Fisica dell'Università di Pisa, Largo B. Pontecorvo 3, 56127 Pisa (Italy)}

\emailAdd{francesco.renga@roma1.infn.it}

\abstract{High precision experiments with muons and pions often require tracking charged particles with $O(\SI{100}{\micro\meter})$ single-hit resolution, possibly with particle identification capabilities, down to very low momenta ($p \lesssim 100$~MeV/$c$). In such conditions, the particle trajectories are strongly affected by the interaction with the detector material, and the reconstruction of the kinematic observables consequently deteriorates. A good compromise between resolution and material budget can be obtained with a Time Projection Chamber (TPC), if very light gases and a high-granularity readout are used. In this paper, we present a characterization of the GridPix detector in helium-isobutane gas mixtures, within a TPC with 9~cm maximum drift. Measurements of the main electron drift properties for these gas mixtures are also presented.}

\keywords{Gaseous detectors, Time projection Chambers (TPC), Micropattern gaseous detectors, Charge transport and multiplication in gas, Particle tracking detectors.}

\arxivnumber{2305.03599} 

\begin{document}
\maketitle
\flushbottom

\section{Introduction}

Experiments with muon and pion beams have been a standard probe for many years to study the Standard Model's flavor structure. Today, cutting-edge searches for new physics beyond the Standard Model are performed at accelerator facilities where a high rate of these particles is stopped in fixed targets~\cite{revip-muon}, or stored with a momentum not exceeding a few hundred MeV/$c$~\cite{g-2,g-2-jp}. Most of these experiments require the reconstruction of the trajectory of either the incoming particle or its charged decay products, whose maximum momentum is, for instance, 52.8~MeV/$c$ for the electrons or positrons emitted in the decay of free muons at rest.

At such low momentum, multiple Coulomb scattering and energy loss affect significantly the trajectory of the particle, such that a precise reconstruction requires an extremely low material budget in the experimental apparatus. Moreover, when these particles move within a $O(1~\mathrm{T})$ magnetic field, curvature radii of $O(10~\mathrm{cm})$ are expected, and the passage of the particle will be measured in a few points (generically called \emph{hits}). It implies that single-hit position resolutions of a few \SI{100}{\micro\meter} will be desirable for a precise measurement of momentum and angles (well below \SI{1}{\percent} and 1~mrad, respectively), by reason of the small lever arms. High granularity is also demanded by the extremely high fluxes arising from the high rate of particles and their confinement in a relatively small volume.

The proposed search for an Electric Dipole Moment (EDM) of the muon at the Paul Scherrer Institut (PSI, Switzerland)~\cite{muEDM} is one example of an experiment having such requirements. For the characterization of the incoming muon beam and the control of the systematic uncertainties on the EDM measurement, the experiment will require reconstruction of the trajectories of 28~MeV/$c$ muons in a 3~T magnetic field, with $O(1~\mathrm{mrad})$ resolution on the longitudinal angle, while exploiting a lever arm of less than 10~cm. In general, low material budget, high resolution, and high granularity are critical requisites for charged-particle tracking in low-energy experiments with pions and muons~\cite{muon-dec,next-meg}.

The proposed solution for the muon beam characterization in the $\mu$EDM experiment is a Time Projection Chamber (TPC) with an extremely light, helium-based gas mixture, read out by a set of GridPix detectors~\cite{gridpix}. Having this application in mind, we present in this paper the first systematic characterization of the GridPix behavior in helium-isobutane gas mixtures, with volume concentrations of 95:5, 90:10, and 85:15. It was necessary to demonstrate the possibility of operating the GridPix with some of the lightest mixtures regularly used in gaseous detectors, and the results will be used as an input to finalize the conceptual design of the detector and assess the reachability of the required performances, which will be the objective of future studies.

The GridPix was installed as the amplification and readout structure of a TPC prototype with a 9~cm maximum drift distance. The measurements were carried out at PSI on the $\pi$M1 beamline of the proton cyclotron accelerator complex, with a 100-150~MeV/$c$ beam composed of positrons ($e^+$, the largely dominant component in this momentum range), pions ($\pi^+$), and muons ($\mu^+$).	

\section{The GridPix}
\label{sec:gridpix}

The GridPix is a gaseous detector made of a Micromegas mesh~\cite{micromegas} mounted on top of a Timepix ASIC~\cite{timepix} for charge collection and readout. The pure aluminum mesh, with \SI{38}{\micro\meter} holes in a grid of \SI{55}{\micro\meter} pitch, is separated from the Timepix by \SI{50}{\micro\meter}-thick
insulating pillars. The microfabrication of the mesh provides a micrometric alignment of the holes on top of the $55 \times 55$~\SI{}{\micro\meter\squared} area pixels of the Timepix. The ASIC is protected from gas discharges by a \SI{4}{\micro\meter}-thick layer of silicon-enriched SiN$_3$.

A voltage difference between the mesh and the Timepix generates the electric field for the multiplication of the electrons originating from the primary ionization and drifting towards the readout structure. The resulting avalanche induces a charge signal in the pixels of the Timepix. The precise alignment of the holes with the pixels ensures that an electron entering one of the holes generates a signal only in the corresponding pixel.

In our application we employ the readout of the Timepix ASIC through the RD51 Scalable Readout System (SRS)~\cite{SRS}, with a dedicated adapter card connected to one front-end concentrator card (FEC)~\cite{FEC}. The acquisition can be externally triggered by an inverted-logic, low-voltage TTL signal. The signal duration determines the Timepix acquisition window. Besides returning the indices of the pixels that registered a signal over a preset threshold within the acquisition window, the ASIC can also provide timing information, operating in two alternative modes: in the time-of-arrival (TOA) mode, the Timepix returns the edge time of the collected signal for each fired pixel, measured backward from the end of the trigger signal; in the time-over-threshold (TOT) mode, the duration of the signal is measured, which is proportional to the induced charge.

\section{The TPC prototype and the beam test setup}

\begin{figure}
\centering 
\begin{subfigure}[b]{0.6\textwidth}
	\centering
	\includegraphics[width=\textwidth]{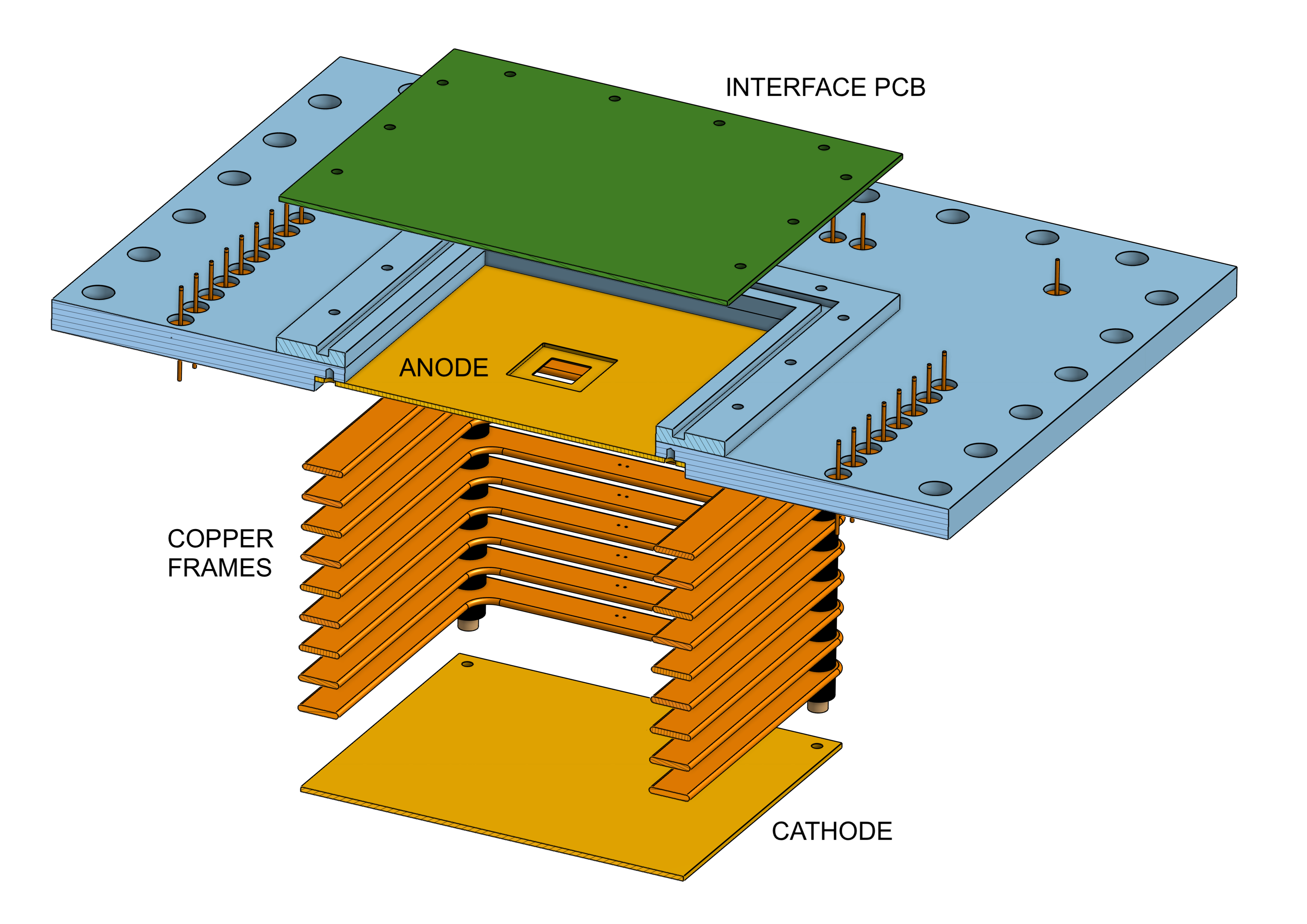}
	\hspace{3cm}
	\caption{}
\end{subfigure}
\begin{subfigure}[b]{0.3\textwidth}
	\centering
	\includegraphics[width=\textwidth]{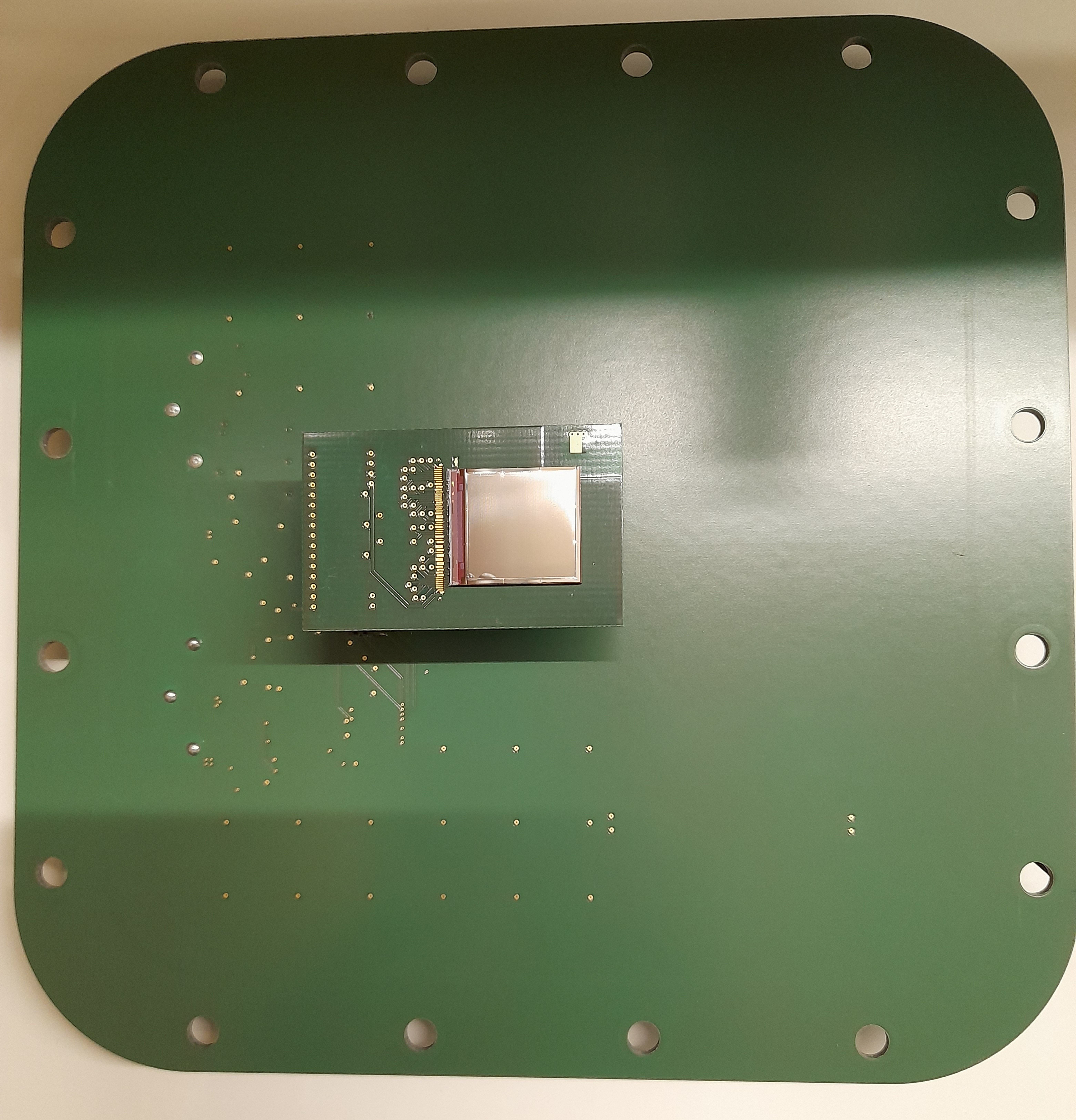}
	\includegraphics[width=\textwidth]{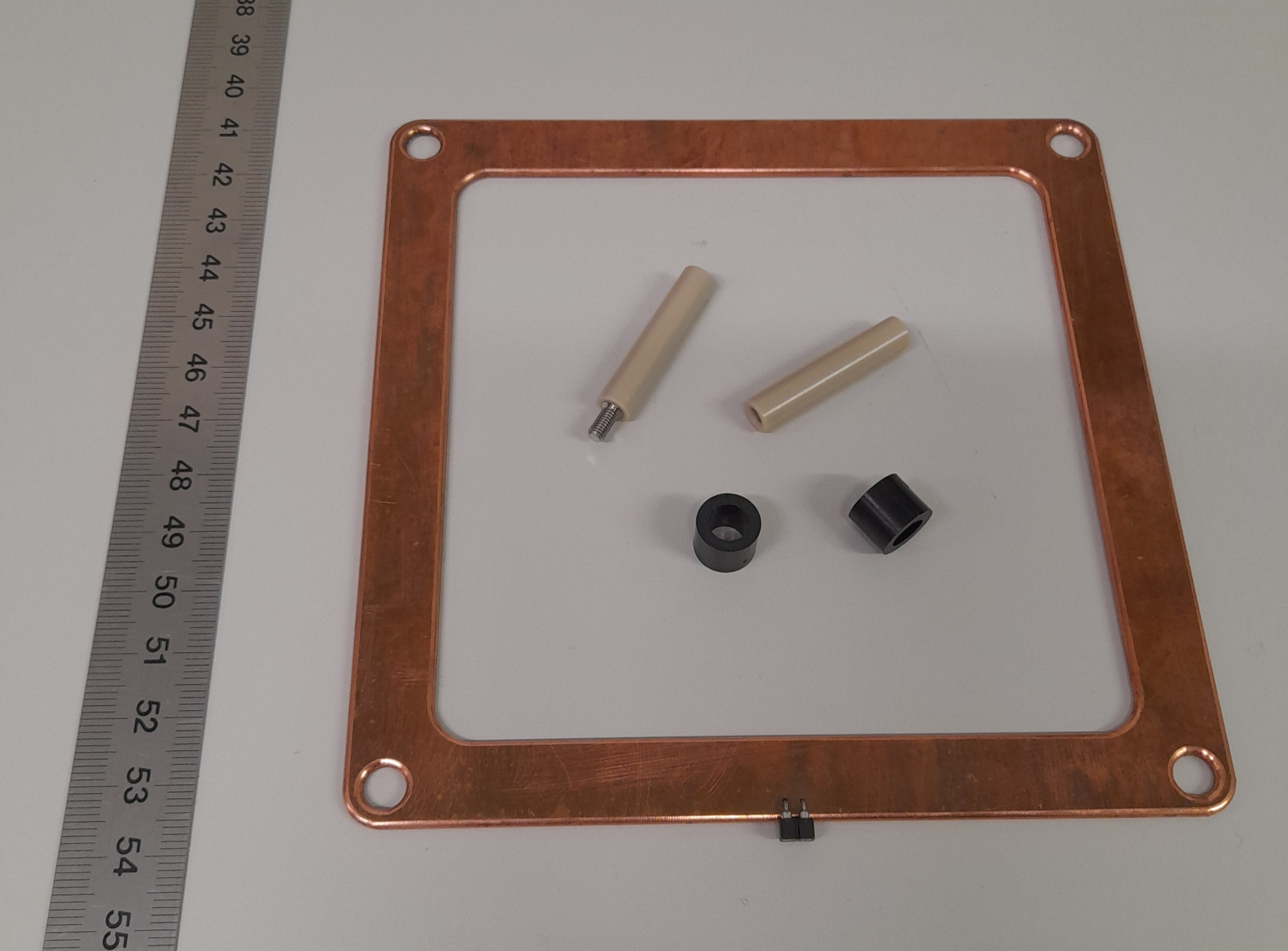}
	\caption{}
\end{subfigure}
\caption{(a) Exploded section view of a simplified model of the internal parts of the detector, mounted on the top face of the PMMA box (in cyan). From top to bottom: the interface PCB (in green); the anode plane with the square hole facing the GridPix; the copper frames; the cathode plane. The Polyamide spacers (in black), skewered on the PEEK pillars (in brown) are also visible. (b) Pictures of the interface PCB with the GridPix sensor (top) and the field cage parts (bottom).}
\label{fig:cage}
\end{figure}

A TPC prototype was constructed and instrumented with one GridPix sensor. A simplified model of the internal components of the detector is shown in figure~\ref{fig:cage}, along with selected images of the parts.

The gas mixture was contained in a Poly(methyl methacrylate) (PMMA) box with windows and flanges sealed with O-rings or flat gaskets. A  window was cut out on each of two opposite lateral faces and each closed by a \SI{25}{\micro\meter} Ethylene tetrafluoroethylene (ETFE) foil. A set of metallic pins and push-in gas fittings were installed in the top face to feed high voltages and provide gas inflow (outflow) to (from) the box. The top face also hosted a $10.4 \times 10.4~\mathrm{cm}^2$ window closed by a PCB to serve as a high voltage, power, and data interface for the GridPix. The sensor itself was bonded on a smaller board, stacked on this PCB. 

Inside the box, a plastic plate was mounted, facing the GridPix. At the center of the plate, a $1.35 \times 1.35~\mathrm{cm}^2$ opening was cut, aligned to the active part of the sensor. The opposite side of the plate was covered by overlapping copper tape strips (\emph{anode plane}). The distance between this metalized surface and the GridPix mesh is $\Delta_\mathrm{anode} = 0.8$~mm. Below the metalized plate, 8 square copper frames, 2~mm thick, with a $10 \times 10~\mathrm{cm}^2$ bore, were stacked with a pitch $\Delta_\mathrm{frame} = 10$~mm. This distance was kept within 0.1~mm tolerance by Polyamide spacers, skewered on Polyether ether ketone (PEEK) pillars going through holes at the corners of the frames. Finally, a glass-fiber plate with a copper deposit (\emph{cathode plane}) was mounted at the end of the stack. Hereafter, this assembly will be referred to as the \emph{field cage}.

The anode, the cathode and the 8 frames provide an approximately constant voltage gradient at the boundaries of a closed region of space, so as to generate a uniform electric field $E_\mathrm{drift}$ above the GridPix. With the Timepix chip at ground voltage and the GridPix mesh at voltage $-V_\mathrm{gain}$, the anode was kept at $-V_\mathrm{anode} = -V_\mathrm{gain} \, - \, \Delta_\mathrm{anode} \cdot E_\mathrm{drift}$, and the cathode was kept at $-V_\mathrm{cathode} = -V_\mathrm{anode} \, - \, 9 \cdot \Delta_\mathrm{frame} \cdot E_\mathrm{drift}$. The first field cage frame with the anode, the last frame with the cathode, and all pairs of consecutive frames were connected through 100~M$\Omega$ resistors with 0.1~\% tolerance. The voltages were generated by a CAEN NDT1470 high-voltage power supply.

The TPC was placed inside a box made of composite polyester-aluminum sheets, serving as a Faraday cage, with shielded feed-throughs for high voltage, power, gas, and GridPix data. Two windows were cut in the box to allow the beam to go through without being perturbed.

A mixture of helium and isobutane was obtained using two mass-flow controllers. Before installing the TPC for the data taking, their relative response was calibrated to 1~\% accuracy against a reference mixture having $(9.8 \pm 0.1)~\%$ isobutane concentration, by means of a Teledyne 7300A infrared gas analyzer. Consequently, the isobutane concentration for the 90:10 mixture, for example, can be quoted as $(10.0 \pm 0.1)~\%$. During the data taking, the mixture was prepared and fed into the chamber with a flow rate of around 12~sl/h, while an electronic flow control valve on the gas outlet regulated the pressure inside the chamber at \SI{0.7}{\milli\bar} above the atmospheric pressure, which was observed to vary between \SI{970}{\milli\bar} and \SI{982}{\milli\bar}.

The specifications of the $\pi$M1 beamline, where the TPC was tested, can be found in~\cite{piM1}. The beam was impinging on a 0.5~mm thick scintillating counter, read out by a Silicon Photomultiplier (SiPM) array, positioned adjacent to the upstream entrance window of the chamber. A 2~cm thick scintillating counter, read out by a Photomultiplier Tube (PMT), was positioned immediately adjacent to the downstream exit window. The two counters would therefore intersect all incoming and outgoing trajectories of particles that could traverse the TPC within the acceptance of the GridPix sensor. The apparatus was approximately aligned to have the beam parallel to the anode plane and approximately aligned above the GridPix sensor. A vertical translation stage allowed the distance between the beam and the GridPix to be adjusted. 


A coincidence of the SiPM and PMT signals was used to trigger a \SI{50}{\micro\second} inverted-logic TTL gate, which was fed into the SRS FEC board to open and close the acquisition window of the GridPix. The average time delay between the SiPM array signal and the end of the acquisition window, equal to \SI{50.851}{\micro\second}, was measured using an oscilloscope and was corrected for the cable lengths. The jitter of the delay amounted to $\approx \SI{50}{\nano\second}$, mostly due to the coincidence logic, which was implemented inside the FPGA of a digitization board\footnote{DRS4 evaluation board, \texttt{https://www.psi.ch/en/drs/evaluation-board}} acquiring the signals from the counters. Most of the data was taken with beam settings giving a coincidence rate of about 300 Hz.

When the charged particles from the beam ionize the gas inside the field cage, the electrons drift in the uniform electric field with a constant velocity $v_\mathrm{drift}$ toward the anode plane and the GridPix, where they can be detected. We alternated the TOT and TOA operation modes of the GridPix to study the response of the sensor and the electron transport properties of the gas mixtures, as described in the next sections.

\section{GridPix efficiency}
\label{sec:gain}

\begin{figure}
\centering 
\begin{subfigure}[b]{0.467\textwidth}
	\centering
	\raisebox{1.25mm}{\includegraphics[trim=0 0 0 -1cm, width=\textwidth]{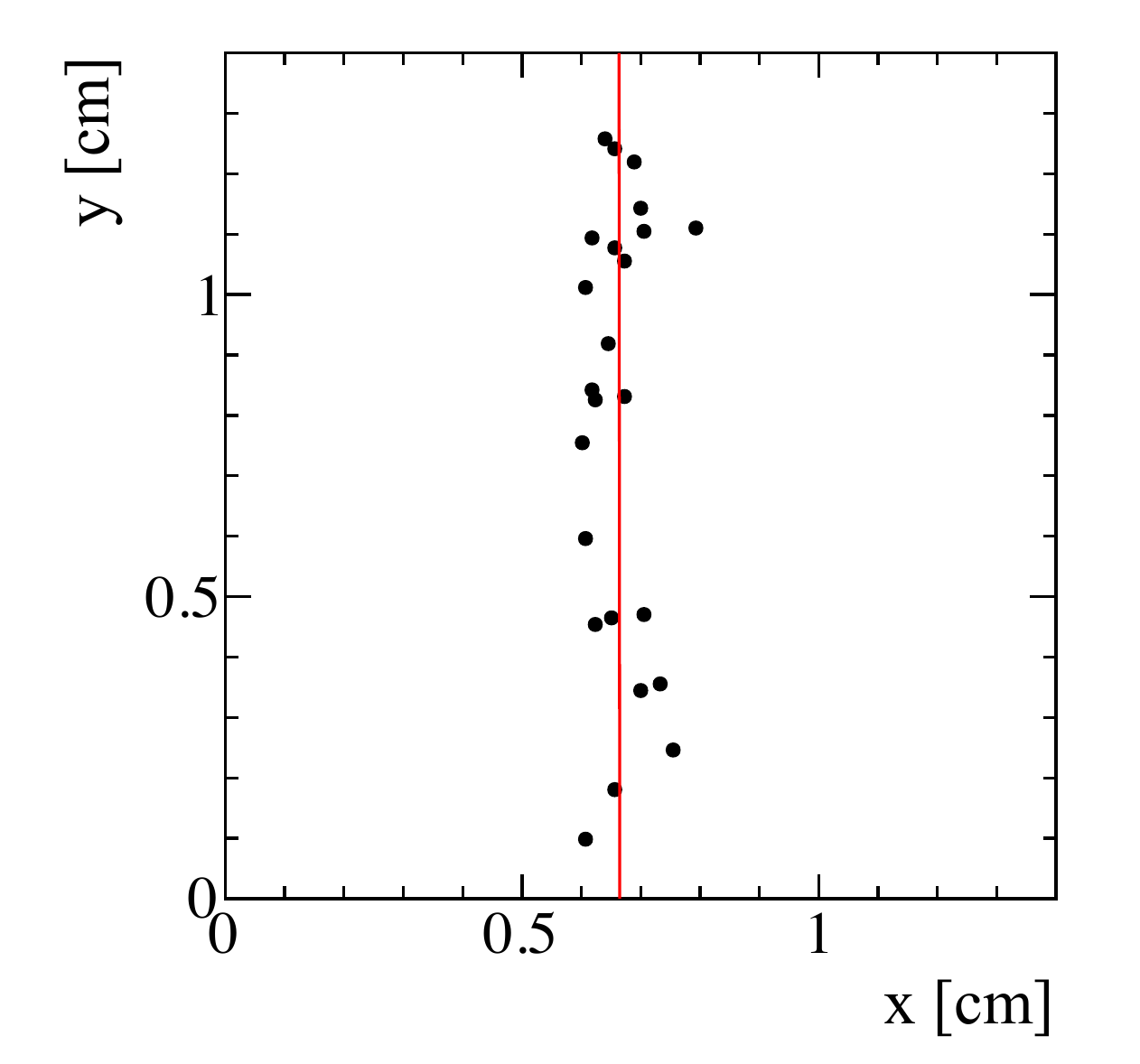}}
	\caption{\label{fig:track2D}}
\end{subfigure}
\begin{subfigure}[b]{0.525\textwidth}
	\centering
	\includegraphics[width=\textwidth]{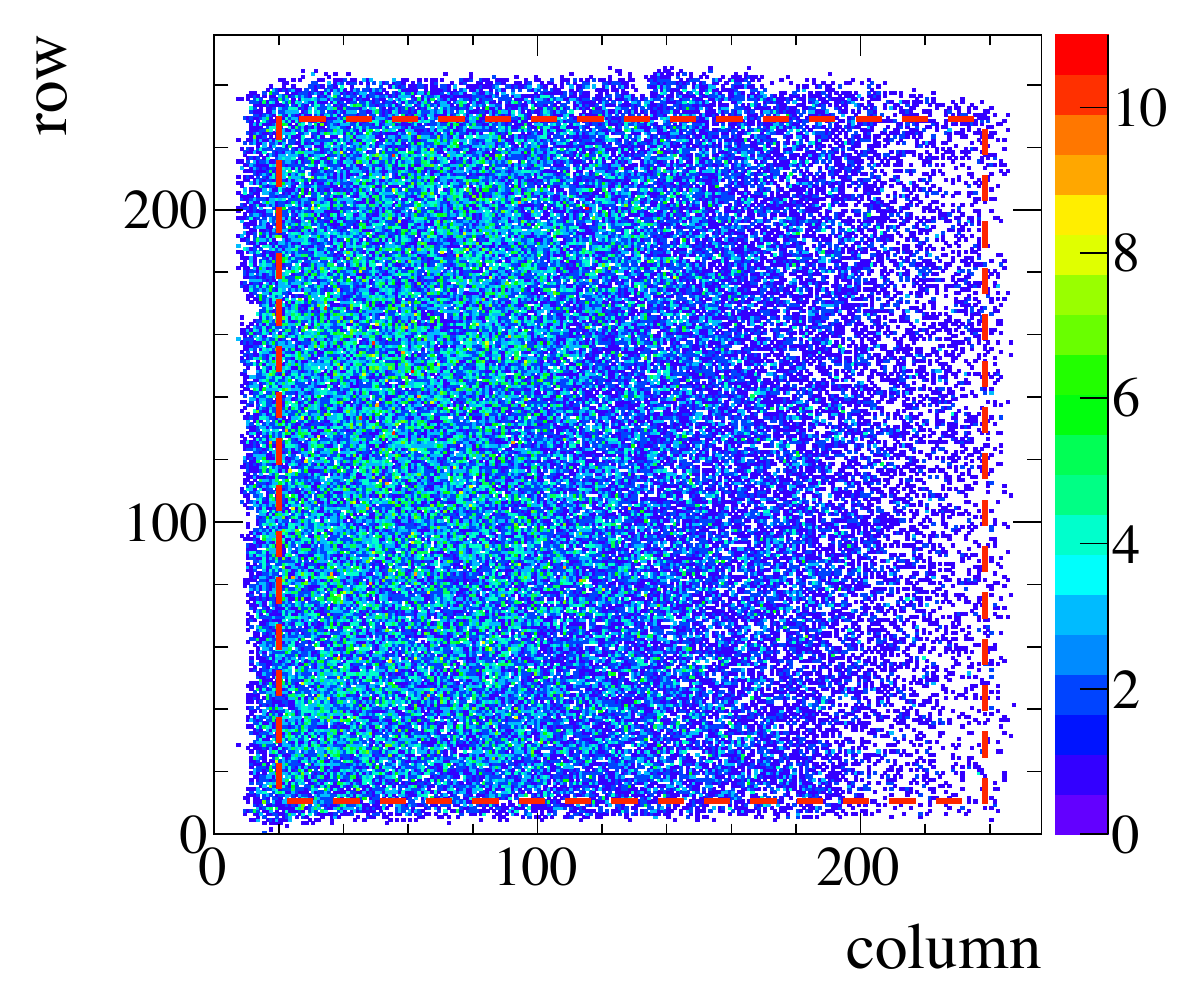}
	\caption{\label{fig:occupancy}}
\end{subfigure}
\caption{(a) An example of a reconstructed 2-dimensional track recorded with the GridPix. (b) Occupancy of the sensor over 5000 trigger events. The trend along the horizontal axis reproduces the shape of the beam spot. The fiducial region is defined as the area inside the red box.}
\end{figure}

In a first step, we determined the minimum operating voltage that must be applied to the GridPix mesh, in order to achieve fully efficient detection of the ionization electrons reaching the sensor. Data were collected in TOT mode, under stable beam conditions, with incremental steps of the mesh voltage $V_\mathrm{gain}$, starting from 300~V. The drift field was fixed to 700 V/cm. Figure~\ref{fig:track2D} shows a track in the GridPix, produced by the passage of a charged particle from the beam. The $y$ coordinate is parallel to the beam direction, the $x$ coordinate is horizontal and orthogonal to the beam axis. In order to identify particle trajectories, a straight-line fit was performed, by minimizing a $\chi^2$ given by the sum of the squared pixel-to-line distances. A resolution of 1~mm was conventionally assumed and, if the reduced $\chi^2$ is lower than 4, the number of hits on the track was counted. To reduce biases induced by the fringe fields at the borders of the sensor, a fiducial region of $220 \times 220$ pixels ($1.21 \times 1.21$~\SI{}{\centi\meter\squared}) cutting out the borders was applied. Figure~\ref{fig:occupancy} shows the occupancy of the sensor over 5000 trigger events and indicates the fiducial region by a red dashed box.

\begin{figure}
\centering 
	\includegraphics[width=0.75\textwidth]{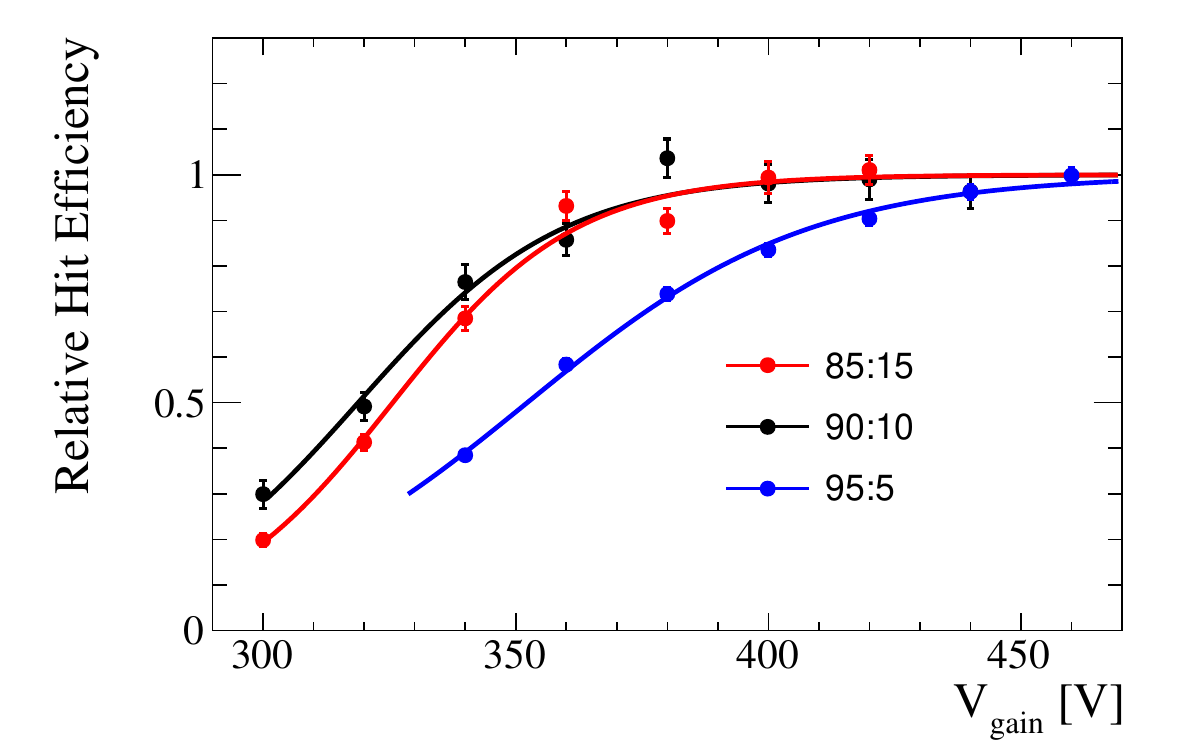}
	\caption{\label{fig:plateau} Average number of hits per event, as a function of the mesh voltage $V_\mathrm{gain}$, normalized to the asymptotic value of the sigmoid curve fitting the data points, for the three different mixtures of helium-isobutane (95:5, 90:10 and 85:15).}
\end{figure}

The average number of hits per event was studied as a function of $V_\mathrm{gain}$, as shown in figure~\ref{fig:plateau}. A fit with a sigmoid function was performed, and the yields were normalized to have the asymptotic value of the curve equal to 1. Above 460~V, the GridPix becomes unstable and all pixels are fired, independent of the gas mixture. 

\begin{figure}
\centering 
	\includegraphics[width=0.75\textwidth]{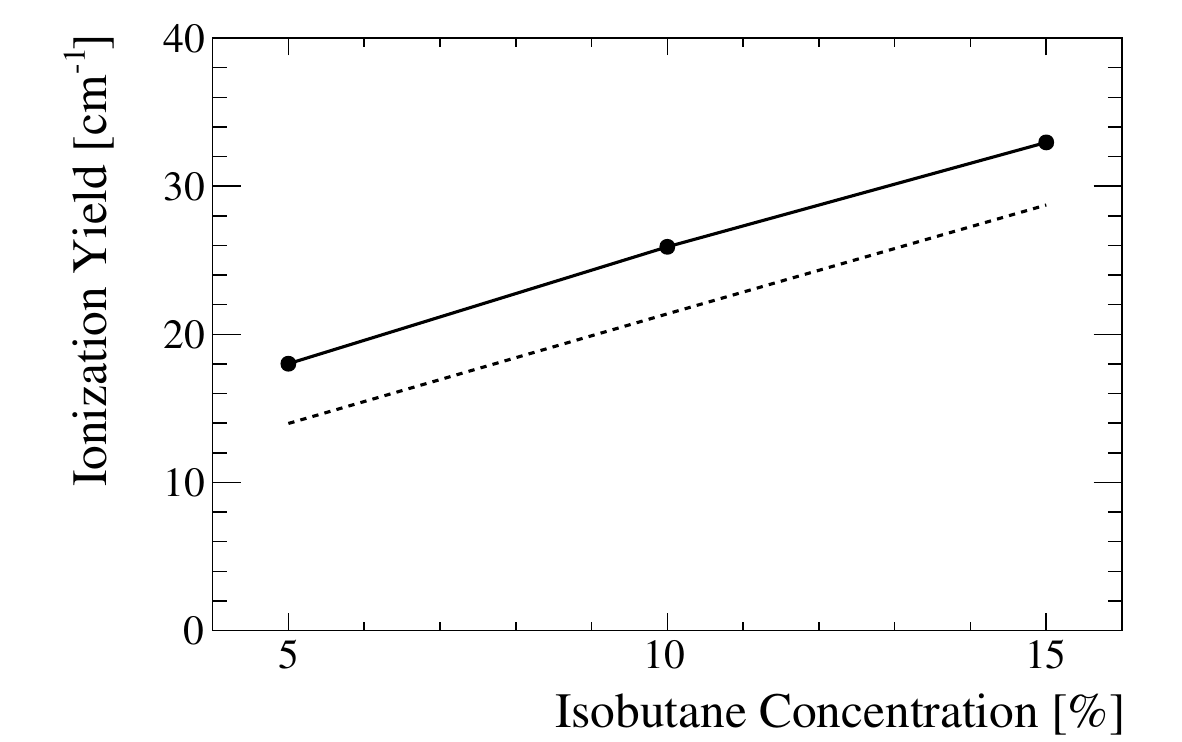}
	\caption{\label{fig:ionization} Average number of hits per unit length (\emph{ionization yield}), as a function of the isobutane concentration, at the selected operating voltages: $V_\mathrm{gain}$ = 440~V (95:5 mixture), 420~V (90:10) and 400~V (85:15). The dashed line shows the GARFIELD++ predictions.}
\end{figure}

Figure~\ref{fig:ionization} shows the average number of hits per unit length, for the three different mixtures at their selected operating voltage. The ionization yield predicted for positrons by GARFIELD++ simulations~\cite{garfield}, also shown in the plot, is systematically lower than the observed one. It can be presumably ascribed to the pion and muon contamination of the beam, along with their higher ionization capabilities according to their larger $dE/dx$, and the absence of Penning ionization~\cite{penning} for helium:isobutane mixtures in GARFIELD++.

The results show that both the 85:15 and the 90:10 mixtures give full efficiency above 400~V, while the efficiency plateau for the 95:5 mixture is only reached around 460~V, where the sensor also starts to become unstable. This naively unexpected behavior, i.e. a higher working point in the least quenched mixture, could be ascribed to the onset of a sizable Penning ionization with higher isobutane concentrations, counteracting the increasing quenching power.

\begin{figure}
\centering 
	\includegraphics[width=\textwidth]{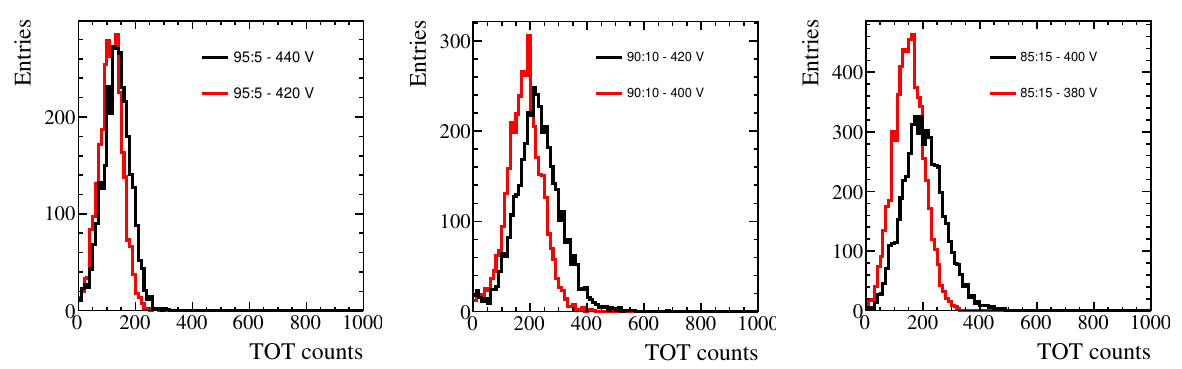}
	\caption{\label{fig:charge} Charge distribution at the operating voltage and 20~V below, for the three helium-isobutane gas mixtures. From left to right: 95:5, 90:10: 85:15. Based on a typical calibration curve for our Timepix, 200 TOT counts correspond to about \SI{2.4}{\femto\coulomb}, which means a gain of about 15000.}
\end{figure}

The data acquired in TOT mode also allowed a study of the distribution of the collected charge per pixel. Figure~\ref{fig:charge} shows the charge distribution per pixel for the three mixtures, in units of the TOT counts, at the operating voltage, and 20~V below.

\section{GridPix charge buildup at high rates}

It is known from previous studies that the GridPix suffers from a loss of gain when exposed to high ionization rates, owing to a negative charge buildup on the protective SiN$_3$ layer, which reduces the electric field in the multiplication gap. In order to study this effect with the different mixtures, the beam aperture was opened to increase the particle rate, which was monitored by the coincidence rate of the two scintillating counters. As the horizontal and vertical dimensions of the counters were approximately the same size as the GridPix and the beam spot, the coincidence rate gave a good estimation of the particle rate in the detector's active volume.

\begin{figure}
\centering 
	\includegraphics[width=0.75\textwidth]{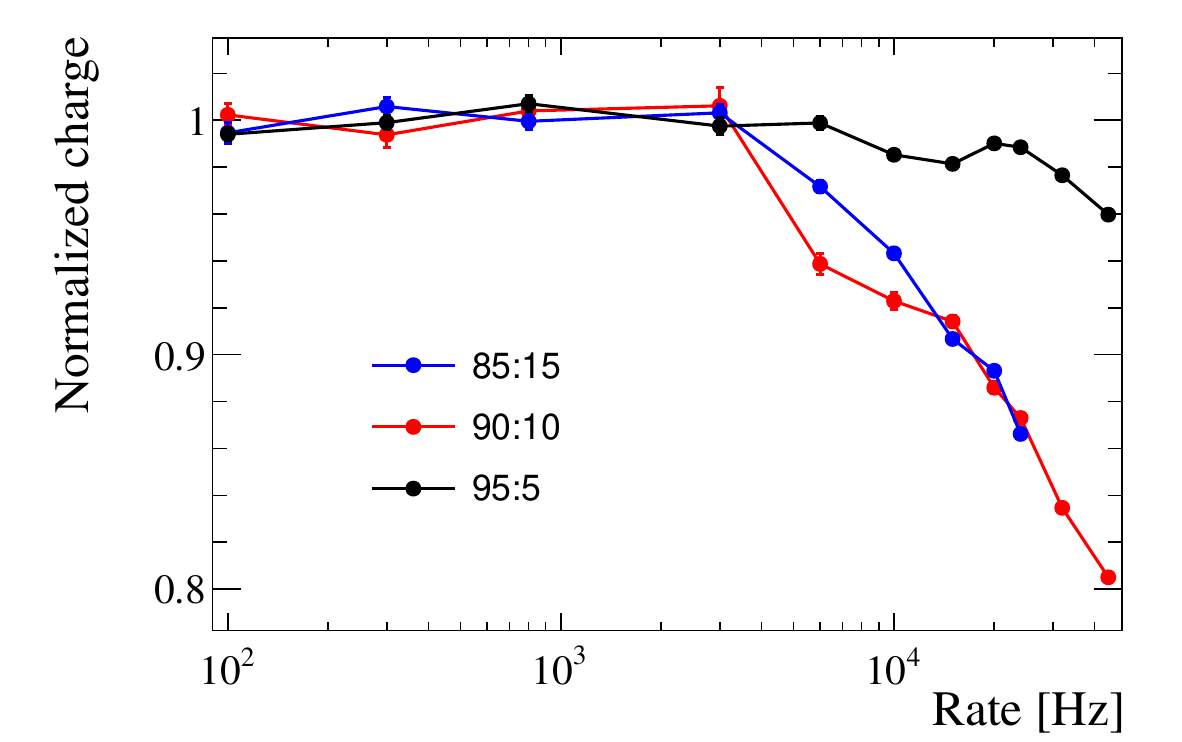}
	\caption{\label{fig:charge_buildup} Average charge per pixel as a function of the beam rate, normalized to the average value of the first three points, for the three different mixtures of helium-isobutane (95:5, 90:10 and 85:15).}
\end{figure}

For this measurement, the GridPix was operated in TOT mode at the operating voltages quoted in the caption of Figure~\ref{fig:ionization}. Figure~\ref{fig:charge_buildup} shows the average charge per pixel as a function of the measured trigger rate, normalized to the average value of the first three points. All mixtures show a significant drop in the average charge, starting around 4 kHz. The drop is less significant in the 95:5 mixture, presumably due to the lower gain at the selected working point (see figure~\ref{fig:charge}). Nevertheless, even in the 85:15 and 90:10 mixtures, the decrease is less than 20~\% around 40 kHz. According to the results presented in section~\ref{sec:gain}, and considering mesh voltages around the working point, it corresponds to a reduction of the mesh voltage of about 15~V, and in turn a loss of about 2~\% of hits. For tracks releasing more than 15 hits per cm, as previously observed, this loss would generate in a detector only a minor deterioration of tracking efficiency and resolutions. Moreover, studies are ongoing to reduce the resistivity of the protection layer in future production cycles, so that accumulated charge will be evacuated more quickly, and it will be possible to operate the sensor at higher particle rates without efficiency loss.

\section{Measurement of electron drift properties}

With a maximum drift length of 9~cm, the prototype also allowed measurement of some drift properties that are relevant for operating a TPC with helium-isobutane mixtures. Having a good understanding of the drift velocity is crucial for various reasons in a detector. Firstly, it is needed to enable a 3-dimensional reconstruction of tracks. Additionally, it plays a role in determining the maximum drift time in the detector, which must be taken into account during the readout design. The spatial resolution of the detector is influenced by the diffusion of the drifting electrons. Lastly, maintaining a low attachment probability is vital to ensure the efficient and accurate operation of the detector.

For these measurements, the GridPix was operated in TOA mode, and different data sets were collected at different vertical positions of the TPC, in order to collect tracks spanning a larger range of distances from the GridPix.

\subsection{Drift velocity}

\begin{figure}
\centering 
	\includegraphics[width=0.75\textwidth]{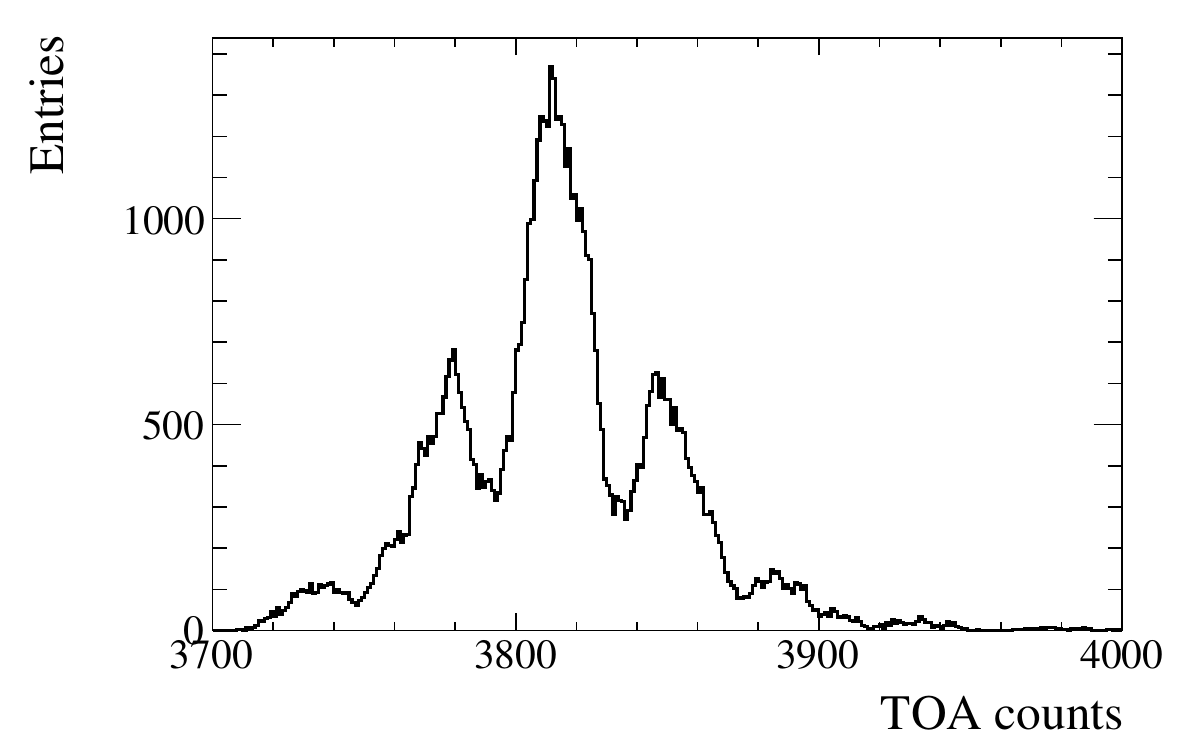}
	\caption{\label{fig:TOA} Distribution of the TOA counts for a vertical position of the TPC that makes the beam spot centered around 6~cm from the GridPix, with a helium-isobutane 90:10 mixture and $E_\mathrm{drift} = 700$~V/cm. A few minima are clearly visible, produced by the copper frames stopping particles at fixed distances, interleaved by 1~cm gaps.}
\end{figure}

Figure~\ref{fig:TOA} shows the distribution of the TOA counts for a given vertical position of the TPC, with the 90:10 mixture and $E_\mathrm{drift} = 700$~V/cm. Each TOA count corresponds to 12.5~ns, and, as mentioned in Sec.~\ref{sec:gridpix}, the time is measured from the leading edge of the signal to the end of the acquisition window, such that higher times correspond to electrons reaching the GridPix earlier. This distribution is the result of ionization by particles passing through the chamber at different distances from the GridPix. The width of the distribution indicates the relatively large beam spot size (about 18 mm Gaussian width), while some minima are produced by the presence of the copper frames of the field cage, which stop particles at specific distances. These features prevent a precise measurement of the drift velocity based on the average drift time as a function of the vertical position of the TPC. However, as the distance between the copper frames is well known, it is possible to associate time differences between minima to drift distances.

\begin{figure}
\centering 
	\includegraphics[width=0.75\textwidth]{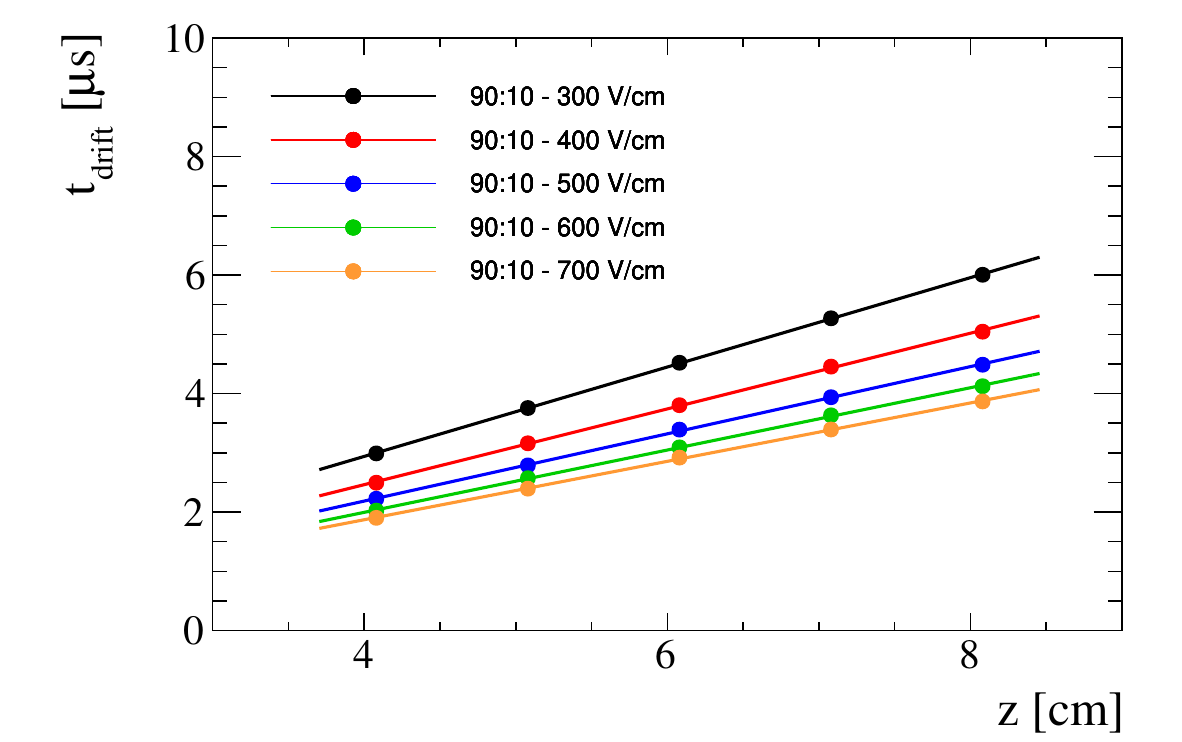}
	\caption{\label{fig:time_distance} Drift times corresponding to the minimum of the minima in the TOA distributions versus the distance of the corresponding frame from the GridPix, for the helium-isobutane 90:10 mixture at different drift fields, with a straight line fit superimposed.}
\end{figure}

\begin{figure}
\centering
	\includegraphics[width=0.75\textwidth]{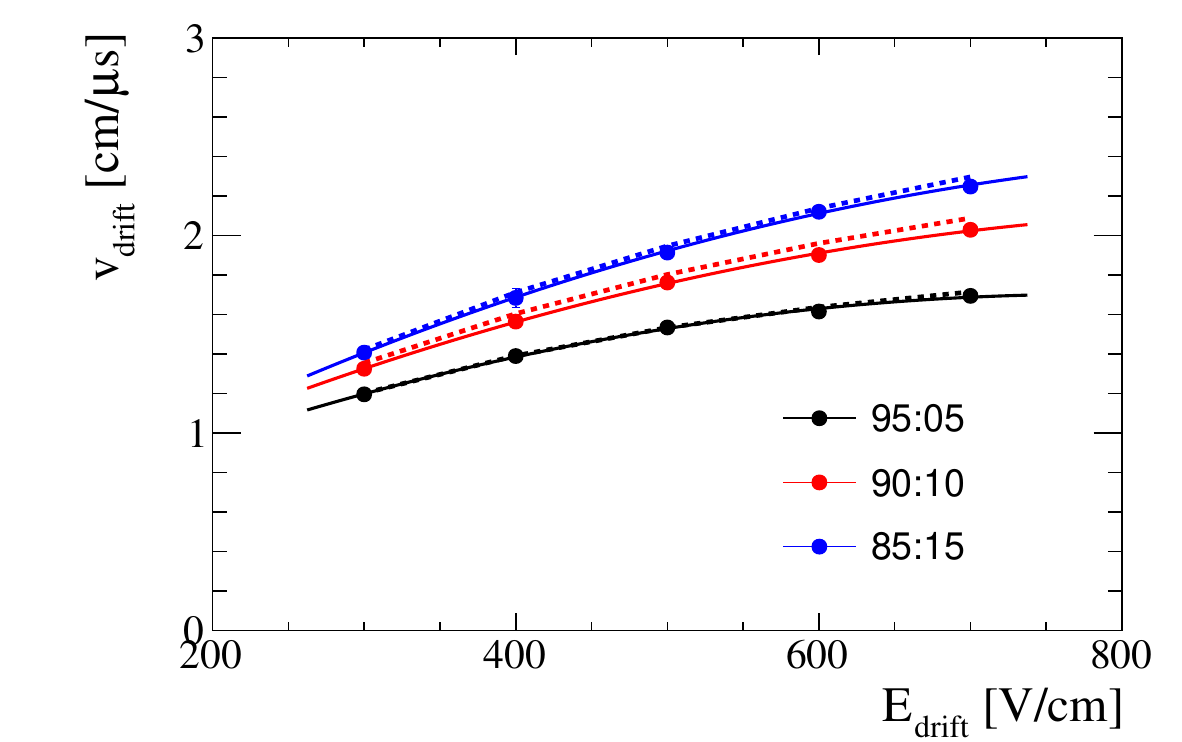}
	\caption{\label{fig:velocity} Measured drift velocity as a function of the drift field for the three different mixtures of helium-isobutane (95:5, 90:10 and 85:15). A fit with a second-order polynomial (full line) and the GARFIELD++ prediction (dashed line) are superimposed.}
\end{figure}

Figure~\ref{fig:time_distance} shows the drift times corresponding to the minima (extracted from second-order polynomial fits around the minimum and corrected for all known time delays\footnote{Once the position of a minimum is known in units of TOT counts, it can be converted into a drift time by $$t_\mathrm{drift} = t_\mathrm{max} - \mathrm{TOT} \cdot \SI{12.5}{\nano\second} \, ,$$ where $t_\mathrm{max}$ is determined with an oscilloscope as the time difference between the end of the acquisition window and the trigger signal, corrected for the cable lengths.}) versus the distance of the corresponding frame from the GridPix, for the 90:10 mixture and different drift fields. Straight-line fits allowed the extraction of the drift velocities, which are depicted in figure~\ref{fig:velocity} as a function of the drift field, for the three different mixtures. The figure also illustrates the good agreement with predictions of the GARFIELD++ simulation software.

\subsection{Diffusion}

\begin{figure}
\centering 
	\includegraphics[width=0.97\textwidth]{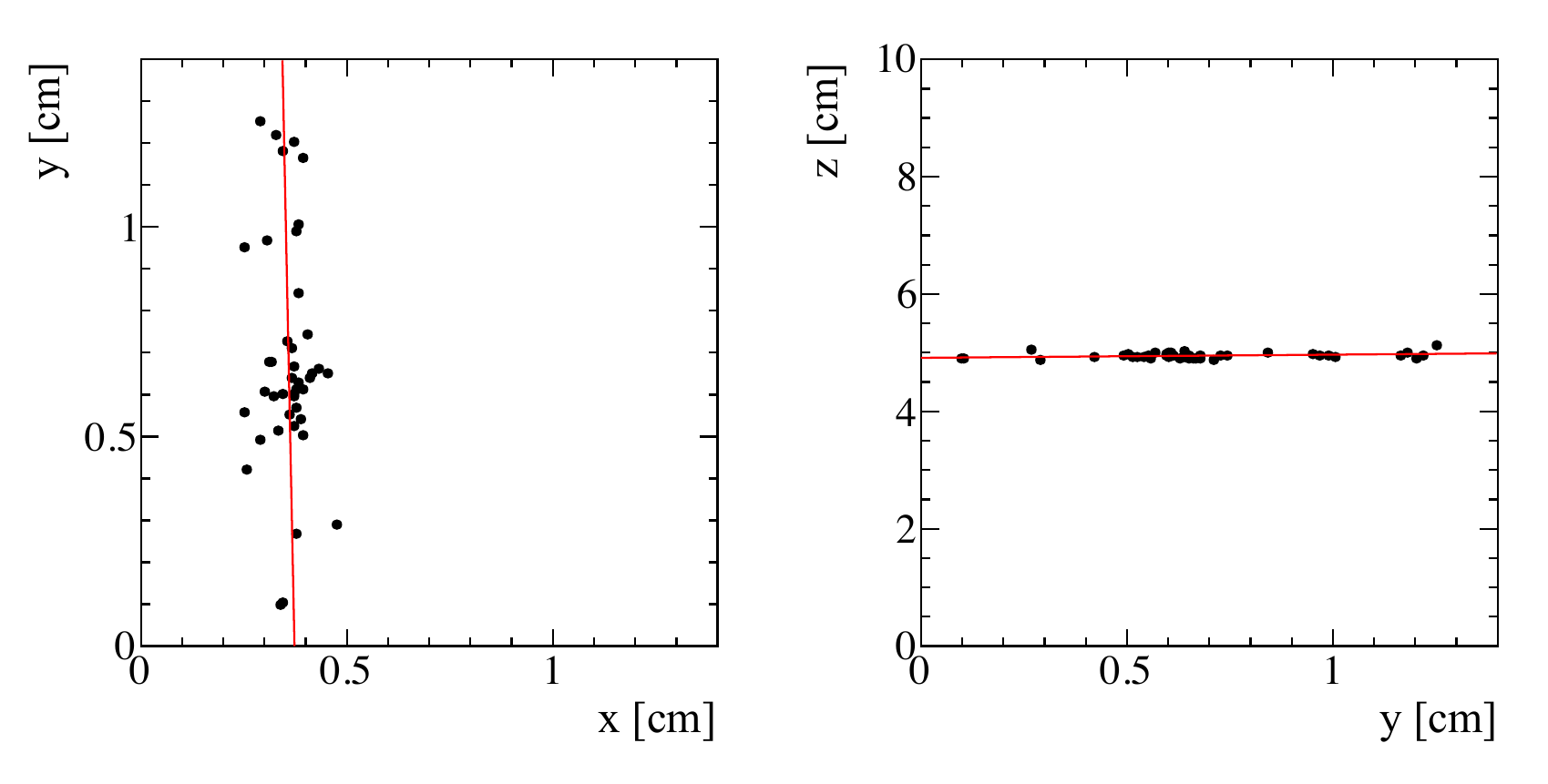}
	\caption{\label{fig:track3D} An example of a reconstructed 3-dimensional track in the GridPix.}
\end{figure}

The measured drift velocities can be used to reconstruct 3-dimensional hits, and hence 3-dimensional tracks, inside the TPC. In this case, the track fit is performed by $\chi^2$ minimization in 3 dimensions. An example is shown in figure~\ref{fig:track3D}, where $z$ is the coordinate along the drift direction (vertical and pointing down).

\begin{figure}
\centering
\begin{subfigure}[b]{0.48\textwidth}
	\centering
	\includegraphics[width=\textwidth]{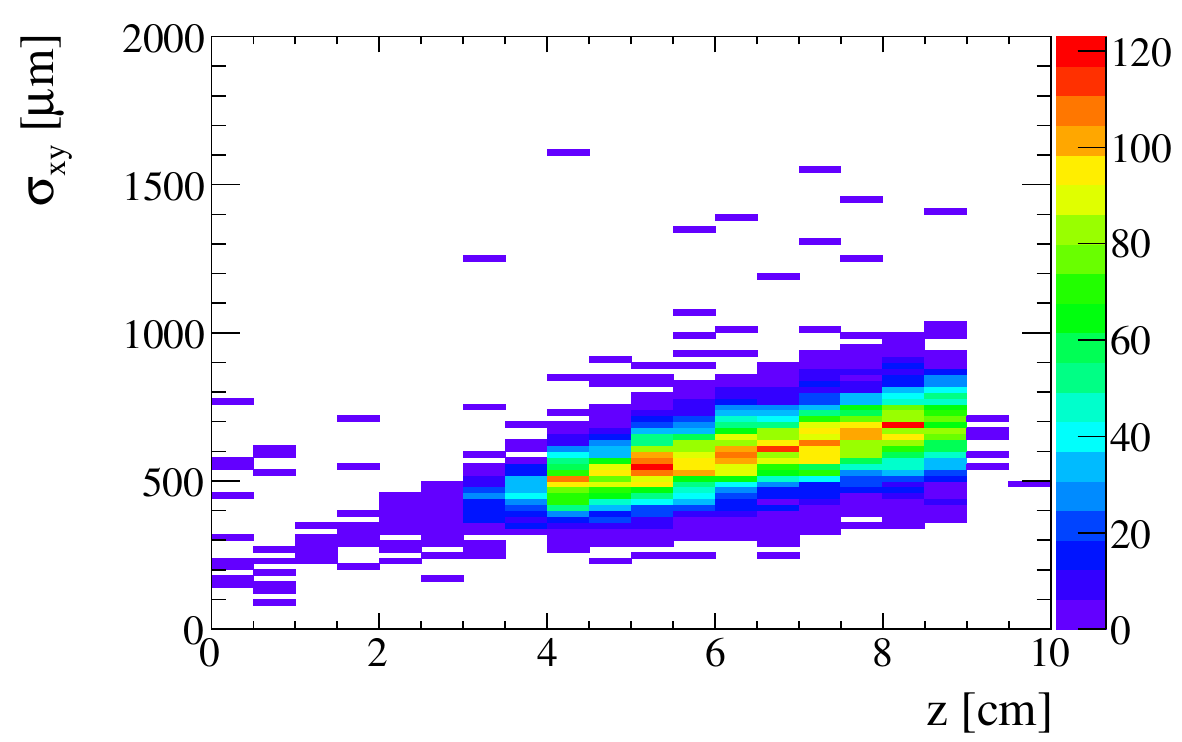}
	\caption{}
\end{subfigure}
\begin{subfigure}[b]{0.48\textwidth}
	\centering
	\includegraphics[width=\textwidth]{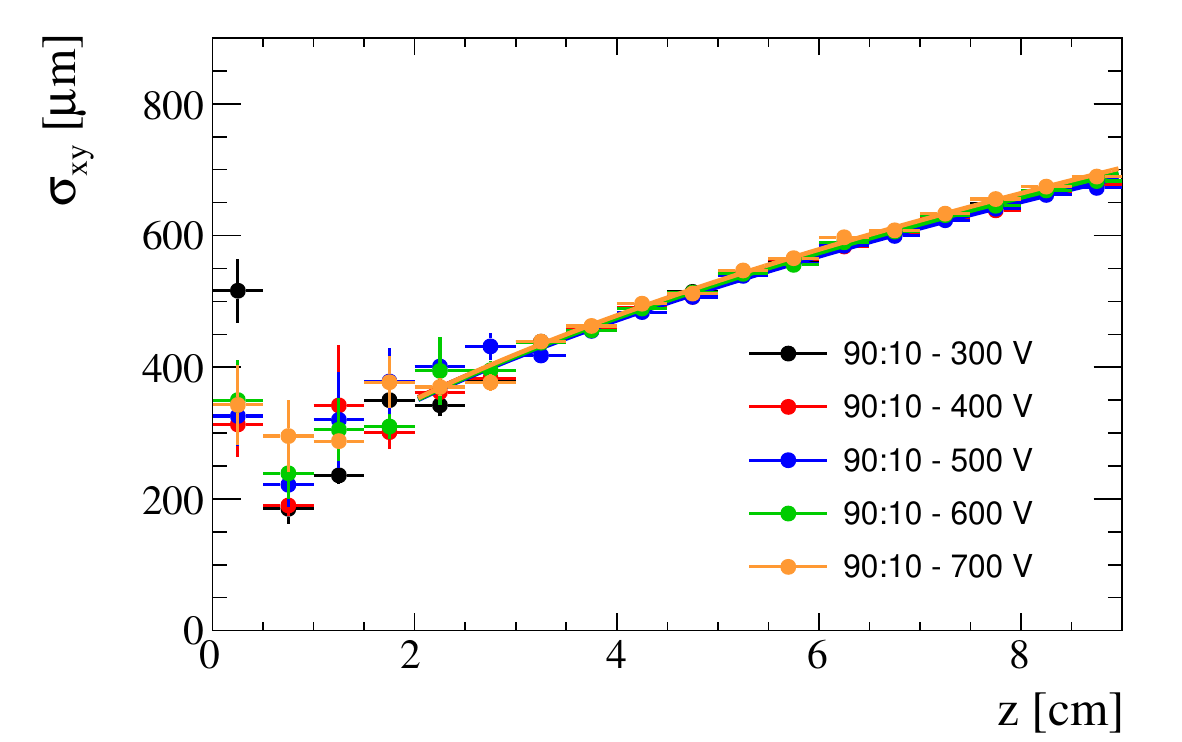}
	\caption{}
\end{subfigure}
\caption{\label{fig:RMS} (a) Distribution of the track-by-track RMS in the $(x,y)$ plane, as a function of the drift distance, for the helium-isobutane 90:10 mixture, at $E_\mathrm{drift} = 700$~V. (b) Average value of the RMS, as a function of the drift distance, for the helium-isobutane 90:10 mixture at different drift fields. See the text for a description of the fit function. Only the points in the range of 2~cm to 9~cm are used in the fit.}
\end{figure}

A longitudinal and transverse diffusion measurement was performed, looking at the RMS of the hit-to-track distances along the drift direction and on the $(x,y)$ plane, respectively. Figure~\ref{fig:RMS} shows the distributions of the track-by-track RMS in the $(x,y)$ plane and their averages, as a function of the drift distance, for the 90:10 mixture at different drift fields. Data are described by the sum in quadrature of the expected $\sqrt{z}$ trend of the diffusion effects and a constant contribution, that can be induced by field imperfections or similar effects:
\begin{eqnarray}
\nonumber \sigma_{xy} &=& \sqrt{(\sigma_{xy}^0)^2 + D_T^2 Z} \, , \\
\sigma_{z} &=& \sqrt{(\sigma_z^0)^2 + D_L^2 Z} \, .
\end{eqnarray}

\begin{figure}
\centering 
\begin{subfigure}[b]{0.48\textwidth}
	\centering
	\includegraphics[width=\textwidth]{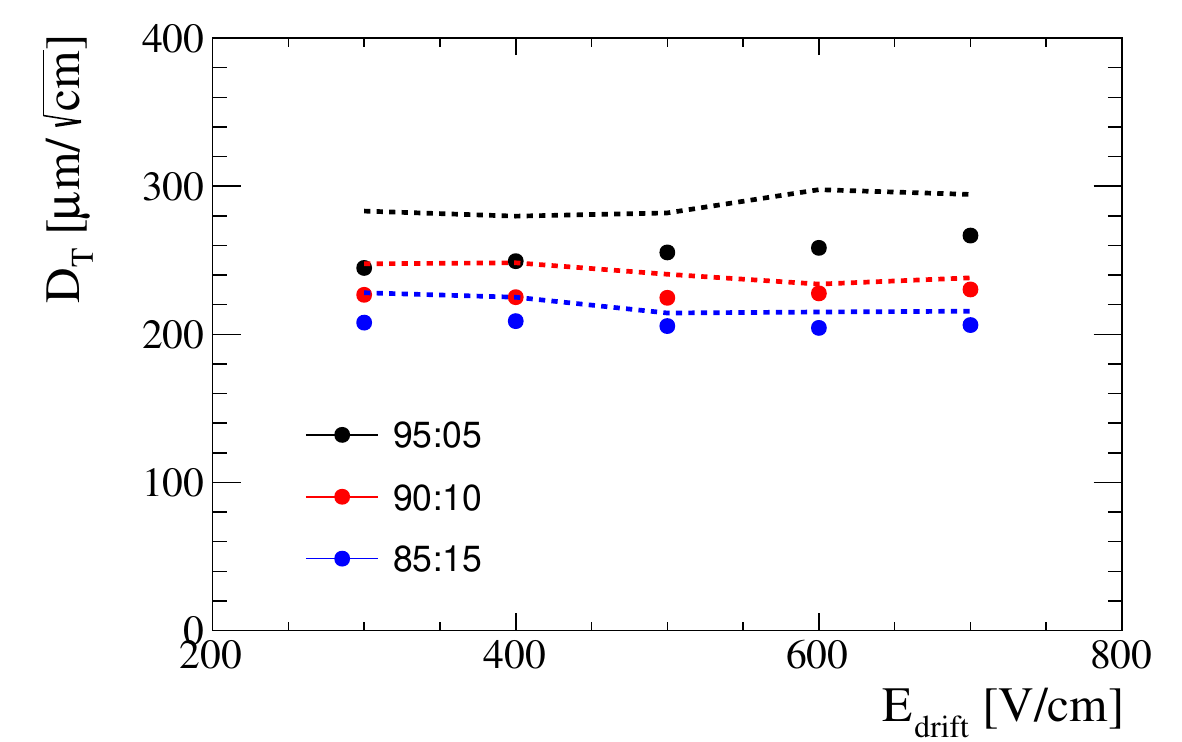}
	\caption{}
\end{subfigure}
\begin{subfigure}[b]{0.48\textwidth}
	\centering
	\includegraphics[width=\textwidth]{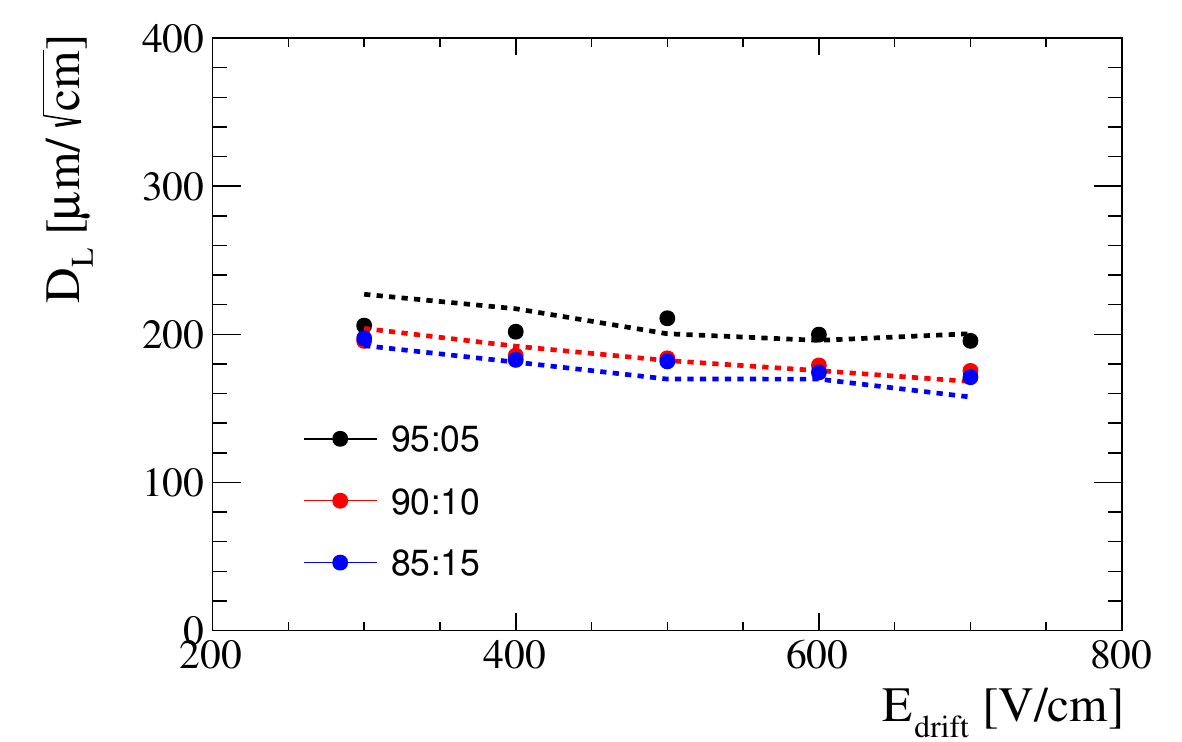}
	\caption{}
\end{subfigure}
\caption{\label{fig:diffusion} Transverse (a) and longitudinal (b) diffusion as a function of the drift field, for the three different mixtures of helium-isobutane (95:5, 90:10 and 85:15) along with the GARFIELD++ predictions (dashed lines).}
\end{figure}

The transverse ($D_T$) and longitudinal ($D_L$) diffusion are shown in figure~\ref{fig:diffusion} as a function of the drift field, for the different mixtures. A qualitative agreement with the GARFIELD++ predictions is found, with discrepancies that probably arise from systematic uncertainties in the simplified gaussian modeling the measurement is based on.

\subsection{Attachment}

Electrons drifting toward the anode can be captured, mainly by electronegative molecules, like oxygen, that can be present as contaminants in the gas mixture. Although contaminations can strongly vary depending on the operating conditions, we measured the attachment coefficient in our beam test setup, as a benchmark for future applications. Indeed, if we can demonstrate that a negligible attachment rate has been achieved in a detector where we did not take any special action to minimize the contaminations, apart from the usual good practices in sealing and choice of detector materials, we can be confident to reach similar or better results in experiments were more efforts could be deployed in this direction.

\begin{figure}
\centering 
\begin{subfigure}[b]{0.48\textwidth}
	\centering
	\includegraphics[width=\textwidth]{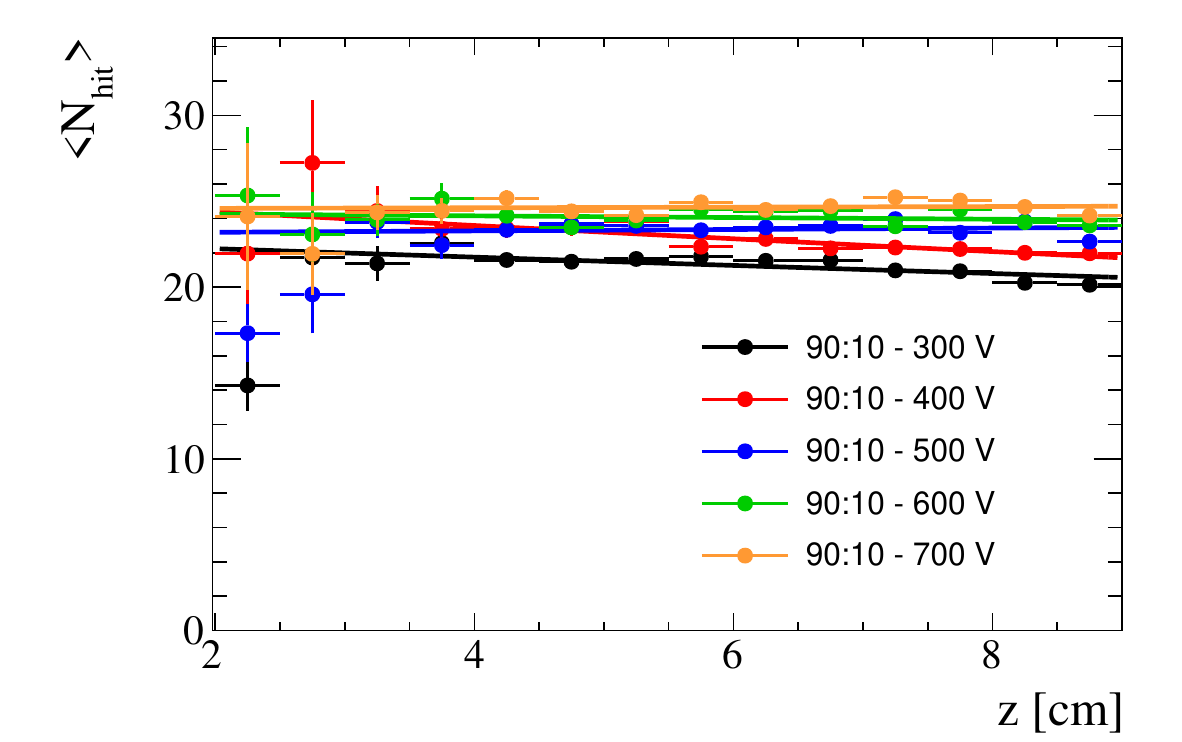}
	\caption{\label{fig:hits_z}}
\end{subfigure}
\begin{subfigure}[b]{0.48\textwidth}
	\centering
	\includegraphics[width=\textwidth]{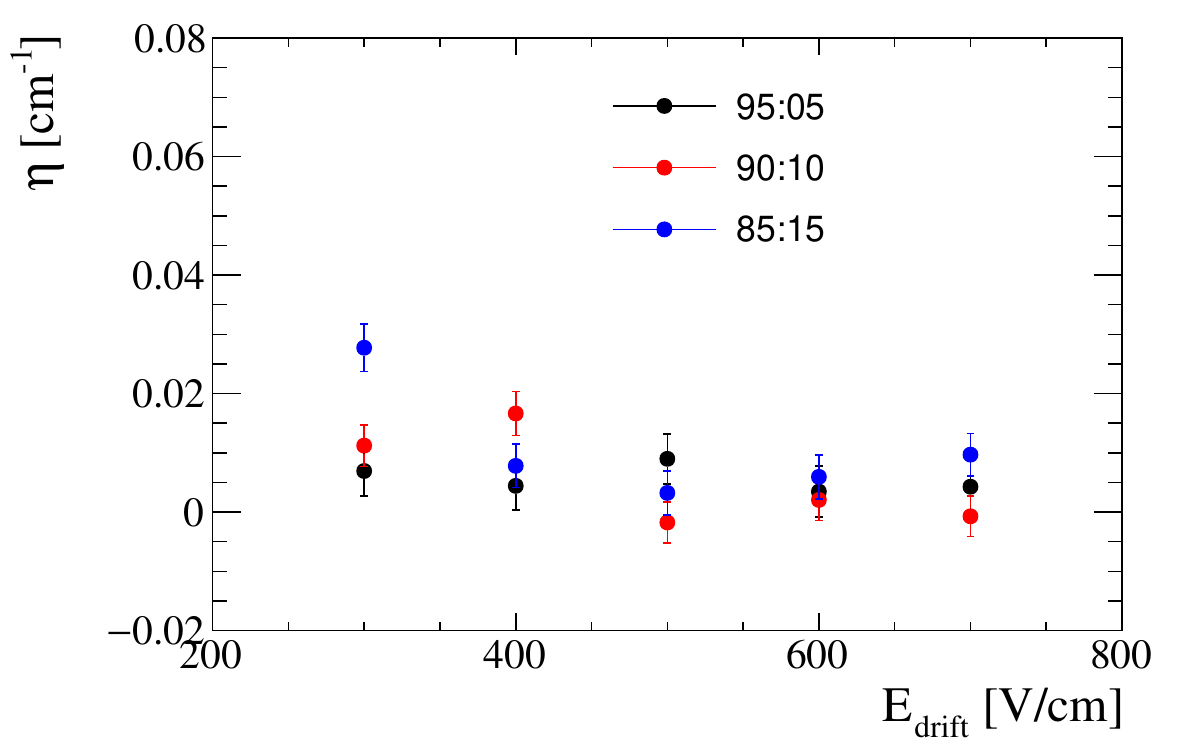}
	\caption{\label{fig:attachment}}
\end{subfigure}
\caption{(a) Average number of hits per unit length as a function of the drift distance, for the helium-isobutane 90:10 mixture at different drift fields, with exponential fits superimposed. Only the points in the range of 2~cm to 9~cm are used in the fit. (b) Attachment coefficient as a function of the drift field, for the three different mixtures of helium-isobutane (95:5, 90:10, and 85:15).}
\end{figure}

The attachment coefficient can be measured by counting the number of hits per cm on selected tracks. Figure~\ref{fig:hits_z} shows the average number of hits per cm, for the 90:10 mixture and different drift fields, as a function of the drift distance. The attachment is expected to reduce the hit yield according to an exponential function:
\begin{equation}
\left<N_\mathrm{hit}\right> = N_0 \, e^{-\eta z} \, ,
\end{equation}
where $\eta$ is the attachment coefficient. Figure~\ref{fig:attachment} shows the measured coefficient for the different mixtures as a function of the drift field.
Although a non-zero attachment coefficient is found with some statistical significance, it is sufficiently small to operate over the full range of drift distances without any relevant effect on tracking.

\section{Conclusions}

In this paper, we have presented a characterization of the GridPix detector with helium-isobutane mixtures, with isobutane concentrations from 5 to 15~\%, in a 9~cm-drift TPC. This study is meant as a demonstration of the possibility of using the GridPix in combination with extremely light gas mixtures for low-momentum particle tracking, which would be especially useful in experiments with pion and muon beams.

Stable operation of the detector was achieved, with 10~\% and 15~\% isobutane concentrations, well above the mesh voltage that guarantees fully efficient detection of ionization electrons reaching the sensor. With 5~\% concentration, higher voltages were needed, and the efficiency plateau was barely reached around 460~V, where the detector starts becoming unstable, independent of the adopted mixture. This naively unexpected higher efficiency of the gas mixture with higher quencher content at lower voltage is tentatively ascribed to the onset of a sizable Penning ionization, more than compensating the quenching effect of higher isobutane concentrations and hence permitting fully efficient operation with lower electric field in the multiplication region.

A positive charge build-up on the SiN$_3$ protection layer was also observed when the detector was exposed to high radiation rates, starting around a few kHz and producing a 20~\% gain loss around a few tens of kHz. In any case, such a loss cannot compromise the tracking efficiency, as long as the detector can be operated at sufficiently high gain so as to remain well within the hit efficiency plateau.

The TPC prototype also allowed a measurement of the electron drift properties of the gas mixtures under study, which were found to be in good agreement with GARFIELD++ simulations. 

Besides their general interest, these results will serve as a background for future simulations, aiming to the finalization of the feasibility studies and to the technical design for a muon tracker in the muon EDM experiment at PSI.

\section*{Acknowledgments} We acknowledge the excellent support provided by the PSI technical groups and by various services of the collaborating universities and research laboratories. This work has received funding from the Swiss State Secretariat for Education, Research and Innovation (SERI), grant No. MB22.00040, and the Swiss National Science Foundation through project No. 204118.

\end{document}